\documentclass[12pt]{spieman}  
\usepackage{amsmath,amsfonts,amssymb}
\usepackage{graphicx}
\usepackage{setspace}
\usepackage{tocloft}
\usepackage{lineno}
\title{Resolving the Origins and Pathways of Ionizing Radiation Escape with UV Integral Field Spectroscopy}

\author[a,b*]{Cody A. Carr}
\author[a,b**]{Renyue Cen}
\author[c]{Brian Fleming}
\author[d]{Sophia Flury}
\author[e]{Stephan McCandliss}
\author[f]{M. S. Oey}
\author[g,h]{Allison Strom}
\affil[a]{Center for Cosmology and Computational Astrophysics, Institute for Advanced Study in Physics, Zhejiang University, Hangzhou 310058,  China}
\affil[b]{Institute of Astronomy, School of Physics, Zhejiang University, Hangzhou 310058,  China}
\affil[c]{Laboratory for Atmospheric and Space Physics, Boulder, CO, USA}
\affil[d]{Institute for Astronomy, University of Edinburgh, Royal Observatory, Edinburgh, EH9 3HJ, United Kingdom}
\affil[e]{Johns Hopkins University, Department of Physics \& Astronomy, Center for Astrophysical Sciences, 3400 North Charles Street,
Baltimore, MD, 21218, United States}
\affil[f]{Astronomy Department, University of Michigan, Ann Arbor, MI 48109, United States}
\affil[g]{Center for Interdisciplinary Exploration and Research in Astrophysics (CIERA), Northwestern University, 1800 Sherman Ave.,
Evanston, IL 60201, United States}
\affil[h]{Department of Physics and Astronomy, Northwestern University, 2145 Sheridan Road, Evanston, IL 60208, United States}

\cftpagenumbersoff{figure}
\cftpagenumbersoff{table} 
\begin{document} 
\maketitle

\begin{abstract}
The Epoch of Reionization marks the last major phase transition in the early Universe, during which
the majority of neutral hydrogen once filling the intergalactic medium was ionized by the first galaxies.
The James Webb Space Telescope (JWST) is now identifying promising galaxy candidates capable of producing
sufficient ionizing photons to drive this transformation. However, the fraction of these photons that escape
into intergalactic space—the escape fraction—remains highly uncertain. Stellar feedback is thought to play
a critical role in carving low-density channels that allow ionizing radiation to escape, but the dominant
mechanisms, their operation, and their connection to observable signatures are not well understood. Local
analogs of high-redshift galaxies offer a powerful alternative for studying these processes, since ionizing
radiation is unobservable at high redshift due to intergalactic absorption. However, current UV space-based
instrumentation lacks the spatial resolution and sensitivity required to fully address this problem. The core
challenge lies in the multiscale nature of LyC escape: ionizing photons are generated on scales of 1–100 pc
in super star clusters but must traverse the circumgalactic medium which can extend beyond 100 kpc.  The proposed Habitable Worlds Observatory (HWO) will provide a platform for future UV instruments capable of resolving these scales. In this article, we present a science case for understanding how LyC photons escape from star-forming galaxies and define the observational requirements for future instruments aboard HWO, including a UV integral field spectrograph (IFS).
\end{abstract}


{\noindent \footnotesize\textbf{*}Cody Carr,  \linkable{codycarr24@gmail.com} \textbf{**}Renyue Cen,  \linkable{renyuecen@zju.edu.cn}}

\begin{spacing}{2}   

\section{Introduction}
\label{sect:intro}  

Approximately one billion years after the Big Bang, the universe experienced its last major phase transition, during which the majority of neutral hydrogen filling the intergalactic medium (IGM) became ionized. This era, known as the Epoch of Reionization (EoR), remains one of the last major frontiers in cosmology \cite{Robertson2022}. Key questions, such as the exact duration of the EoR and the identity of the astrophysical objects responsible for reionizing the universe, remain unanswered. 

Early results from the James Webb Space Telescope (JWST) suggest that star-forming (SF) galaxies produce enough ionizing, or Lyman continuum (LyC; $< 912 $ \AA), photons to reionize the universe \cite{Atek2024,Lin2024,Munoz2024,Pahl2025}. However, it remains unclear how many of these photons actually escape from the first galaxies to ionize the IGM. Neutral gas and dust within the interstellar (ISM) and circumgalactic medium (CGM) efficiently absorb LyC photons, making escape unlikely without some form of intervention. Feedback processes—such as radiation pressure, stellar winds, and supernovae—are thought to create low-density channels that allow LyC photons to leak out. Yet, the dominant mechanism, how these processes operate, and how they connect to observable signatures are still open questions. 

Deciphering the physical mechanisms that govern LyC escape is inherently a multiscale problem—spanning the production of LyC photons in super star clusters (SSCs; $\sim$1–100 pc) to their eventual escape through the circumgalactic medium (CGM; $\sim$1–100 kpc). Currently, the astronomical community lacks space-based instrumentation capable of resolving ultraviolet (UV) emission on SSC scales. This limitation hinders our ability to pinpoint which type of feedback primarily facilitates LyC escape and how it links to various observables. The advent of a 6-meter-class, space-based observatory recommended by the 2020 Decadal Survey \cite{nasem2023}—the Habitable Worlds Observatory (HWO)—will provide the light-gathering power required to enable a UV integral field spectrograph (IFS) capable of spatially resolving super star clusters (SSCs) and kinematically resolving their galactic winds at high signal-to-noise. The goal of this paper is to define the spectroscopic requirements imposed by this science case, whose demanding nature will naturally support a broader range of high-impact UV science programs.

An earlier version of this work appeared as a Science Case Development Document (SCDD) in the conference proceedings of the “Towards the Habitable Worlds Observatory: Visionary Science and Transformational Technology” 2025 summer meeting, published by the Astronomical Society of the Pacific.

\section{Science Case: Linking Stellar Feedback to LyC Escape and UV Observables}

The primary challenge to measuring a galaxy’s contribution to reionization lies in determining what fraction of its intrinsic LyC production actually goes on to ionize the IGM.  This quantity is referred to as the LyC escape fraction ($f_{\mathrm{esc}}^{\mathrm{LyC}}$).  Absorption in the neutral IGM prevents direct measurements of $f_{\mathrm{esc}}^{\mathrm{LyC}}$ at high redshifts \cite{Inoue2014}.  Consequently, astronomers must rely on local analogs of high redshift galaxies and simulations to develop diagnostics of $f_{\mathrm{esc}}^{\mathrm{LyC}}$ to infer the contributions of individual galaxies to reionization \cite{Flury2022b,Choustikov2024,Jaskot2024a,Jaskot2024b}.  For instance, high [O III] 5700\AA/ [O II] 3726,9\AA\ ratios (O32; \cite{Jaskot2013,Nakajima2014,Izotov2017,Izotov2018b,Izotov2020}) and steep non-ionizing UV continuum slopes ($\beta_{\mathrm{UV}}$; \cite{Zackrisson2013,Zackrisson2017,Chisholm2022}) have all been linked to $f_{\mathrm{esc}}^{\mathrm{LyC}}$ in local galaxies.  These quantities contain information about the ionizing intensity and dust content, respectively.  

\begin{figure*}[t]
\begin{center}
\includegraphics[width=\textwidth]{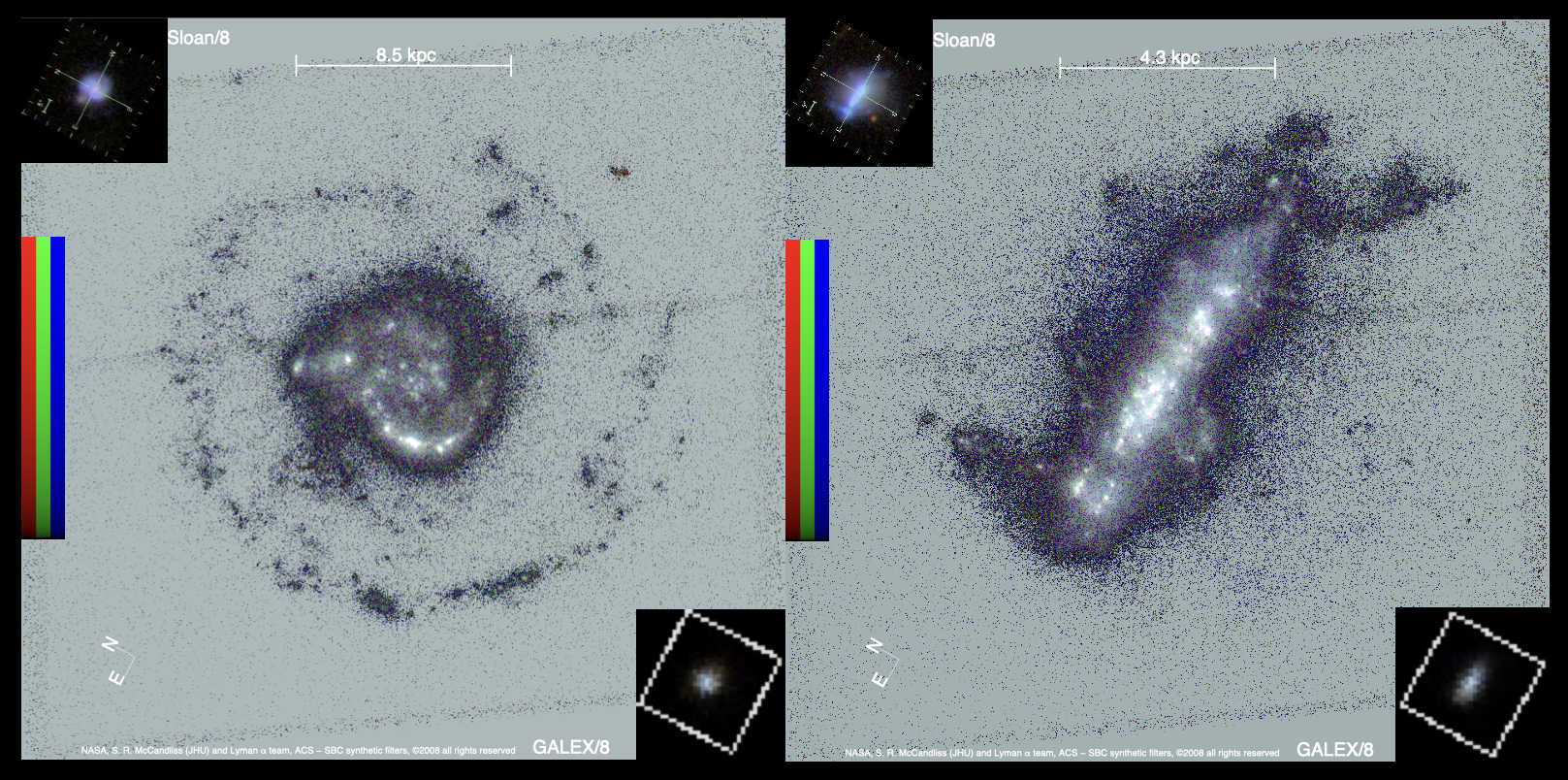}
\caption{ACS/SBC synthetic filter images of two nearby SF galaxies \cite{McCandliss2007}.  Blue subplots show regions of diffuse Ly$\alpha$ emission.  We count roughly 20 (left panel) and 30 (right panel) distinct star clusters in each galaxy.  Spatially resolving such features will be critical for understanding how LyC photons escape from galaxies and linking them to observational signatures.  
\label{fig:clusters}
}
\end{center}
\end{figure*}

Most studies investigating LyC escape diagnostics rely on spatially integrated spectra, such as those obtained with the Cosmic Origins Spectrograph (COS) aboard the Hubble Space Telescope (HST), and therefore cannot capture spatial variations within galaxies. In reality, LyC production is highly localized, occurring primarily within super star clusters (SSCs) composed of massive O- and B-type stars—the dominant sources of ionizing radiation. Figure~\ref{fig:clusters} demonstrates this localization, showing clustered star formation in UV images of two nearby SF galaxies \cite{McCandliss2007}.

\begin{figure*}[t]
\begin{center}
\includegraphics[width=\textwidth]{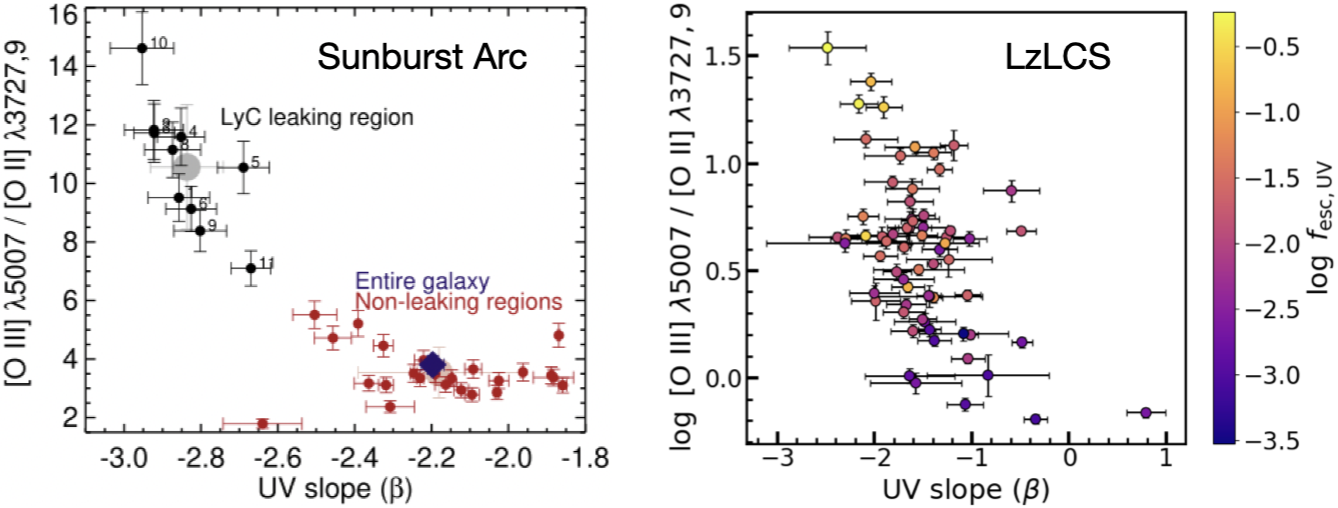}
\caption{LyC escape according to the [O III] 5007\AA\ / [O II] 3726,9\AA\ ratio and the slope of the FUV continuum slope ($\beta$).  The left panel shows values drawn from individual regions of the Sunburst Arc galaxy.  Figure taken from \cite{Kim2023}, with permission.  The black points correspond to regions with LyC emission and the red points without.  The purple point represents the average value measured over the whole galaxy.  The left panel shows values derived from HST COS integrated spectra for entire galaxies selected from LzLCS.  Data were taken from \cite{Flury2022a,Flury2022_erratum}.  To understand the relationship between $f_{\mathrm{esc}}^{\mathrm{LyC}}$ and observables, we must resolve galaxies on the cluster scale.
\label{fig:diagnostics}
}
\end{center}
\end{figure*}

The Sunburst Arc, a strongly lensed LyC-emitting galaxy, offers a rare opportunity to study LyC escape at sub-galactic scales, approaching the level of individual clusters. The right panel of Figure~\ref{fig:diagnostics} illustrates how LyC emission in the Sunburst Arc varies significantly across its surface along with the $\beta_{UV}$ and the O32 ratio \cite{Kim2023}. The diagnostics differ substantially between LyC-emitting (black) and non-emitting regions (red), with O32 generally increasing in regions showing LyC emission.  However, this physical connection between O32 and $f_{\mathrm{esc}}^{\mathrm{LyC}}$ is lost when averaging over the entire galaxy, as is done in integrated spectroscopy (purple).  

For comparison, the right panel of Figure~\ref{fig:diagnostics} displays 66 local galaxies drawn from the Low-z Lyman Continuum Survey (LzLCS; \cite{Flury2022a}) according to these same diagnostics measured from integrated spectra. The resulting distribution shows substantial scatter—particularly at intermediate values of O32 and $\beta_{UV}$, highlighting the loss of information.  To meaningfully connect LyC escape with observable diagnostics—and to uncover the mechanisms underlying these connections—analyzes must be carried out at the scale of individual stellar clusters.

For LyC photons to escape from galaxies into the IGM, SSCs must clear their immediate surroundings and CGM of neutral hydrogen and dust. This clearing is driven by various feedback processes, including radiation pressure, stellar winds, and supernova explosions. Ram pressure from supernova-driven blast waves, for example, can launch massive, multiphase outflows that lift LyC-blocking clouds from the ISM \cite{Borthakur2014,Kimm2014,Cen2020}, while intense ionizing radiation from young stars may create escape pathways by ionizing low-column-density channels which naturally occur in turbulent media \cite{Kakiichi2021,Menon2025}. The dominant feedback mechanism operating in a given region can be inferred from the age of the stellar populations—for instance, radiation feedback typically dominates in clusters younger than 3 Myr \cite{Leitherer1999,Eldridge2017}.  Current observational evidence suggests that LyC escape is highest during the earliest phases of star formation, when young, massive stars emit strong ionizing radiation prior to the onset of supernovae \cite{Hayes2023b,Bait2024,Flury2025,Carr2025LyC,Carr2025Sup}.  In contrast, simulations often favor later evolutionary stages, when supernova-driven winds have had sufficient time to evacuate dense, LyC-blocking gas \cite{Kimm2014,Cen2020,Rosdahl2022,Choustikov2024}, where the LyC production is possibly dominated by binary stars \cite{Secunda2020}.  Disentangling these mechanisms will require spectral imaging of the winds, down to the scale of the SSCs.   
 
The kinematic and geometric (density + distribution) properties of the different outflow phases needed to understand how winds shape their environments is reflected in the broadening of absorption \cite{Prochaska2011,Scarlata2015,Zhu2015,Chisholm2016b,Carr2018} and emission lines \cite{Flury2023,Yuan2023,Amorin2024,Xu2025}.  While empirical measures such as line widths can be recovered from low-resolution spectra (R $<$ 5000), more nuanced parameters—such as density and velocity gradients—require high spectral resolution (R $\sim 15000$, or $\sim$ 20 $\mathrm{km\,s^{-1}}$) for effective radiative transfer modeling \cite{Carr2023,Carr2025MOR}.  However, spatially resolved spectra such as that obtainable with an IFS provides the most promising data set for constraining radial profiles \cite{Burchett2021,Erb2023,Carr2025MOR}.  Furthermore, mapping the winds on a cluster-by-cluster basis, as possible with both a multi-object spectrograph (MOS) or IFS would allow one to pair the feedback mechanism to the locations of LyC escape.  Together, these advances would enable a definitive identification of the physical mechanisms driving LyC escape in the local universe and establish robust connections to LyC escape diagnostics observable at high-redshifts, thereby clarifying the contribution of SF galaxies to reionization.


\section{Desired Observations and Current Limitations}
To understand how LyC radiation escapes from galaxies, we must simultaneously map the H I and dust content of both the ISM and CGM, identify individual stellar populations, and constrain the kinematic and geometric properties of galactic winds. These goals can be achieved through the abundance of UV metal absorption lines observable at low redshifts, which trace a temperature range of roughly $10^4$–$10^6$ K \cite{Tumlinson2017}. In this section, we review the relevant observations and the resulting requirements imposed by our science case. All subsequent claims assume a standard flat cosmology with $H_0 = 70$ km s${}^{-1}$ Mpc${}^{-1}$, $\Omega_m = 0.3$, and $\Omega_\Lambda = 0.7$.

\subsection{Lyman Continuum}

The LyC escape fraction can be measured directly from emission at wavelengths below the 912 \AA\ Lyman limit.  For instance, an empirical proxy is commonly used, where the flux near 900 \AA\ is compared to the continuum flux around 1100~\AA\ \cite{Wang2019,Flury2022a,Jaskot2025}.  Current constraints on $f_{\mathrm{esc}}^{\mathrm{LyC}}$ have been measured using low-resolution spectroscopy ($\rm R \sim 1050$ at 1100 \AA) with the COS G140L grating, which corresponds to an observed wavelength of 900 \AA\ at $z = 0.22$, typically at a signal-to-noise ratio of $\rm S/N = 5$ \cite{Flury2022a}.  However, more recent studies indicate that the estimates $f_{\mathrm{esc}}^{\mathrm{LyC}}$ can change significantly at $\rm R\sim5000$  \cite{Carr2025Sup}.  The COS throughput sensitivity drops by roughly two-orders of magnitude below 1100 \AA, imposing a lower limit at this redshift \cite{Green2012}.  Because we desire to spatially resolve LyC emission on the SSC scale (1-100 pc), the closer we can observe an object the better when observing at low redshifts.  However, contamination from Milky Way absorption will become more and more problematic the closer we observe.  There is an abundance of absorption lines that appear redward of the Lyman limit (e.g., L$\beta$, L$\gamma$, etc.) making clean observations of the LyC difficult.  Realistically, $z=0.16$ is the lowest redshift at which we can comfortably observe LyC emission.  At this redshift, the 900 \AA\ target will be redshifted to 1044 \AA\ in the observed frame.  
 
While most studies focus on stellar contributions to the LyC, in reality, there may be contributions coming from nebular emission as well - for instance, free electrons recombining to the first excited state of hydrogen will create a flux excess just below the Lyman limit at 912 \AA, often called the “Lyman bump” \cite{Inoue2010}.  Since the nebular contribution is anticipated to be highly wavelength dependent, it may influence the perceived value for $f_{\mathrm{esc}}^{\mathrm{LyC}}$.  To ensure separation from the nebular contribution, a broader wavelength range of 700–912 \AA\ would be ideal, as the 700 \AA\ mark is likely far enough away from the nebular contributions from higher order H I transitions \cite{Simmonds2024}.  If we extend to even bluer wavelengths, down to 400 \AA, we would also be able to measure the He Lyman bump below its Lyman limit at 504 \AA.  This must be done at much higher redshifts, however.  For example, 400 \AA\ would be observed at 1100 \AA\ at $z = 1.75$.  Reports of possible LyC detections have been made using COS at $\rm R = 1250-2400$ at 1500 \AA\ over the restframe wavelength range 770 - 912 \AA\ \cite{Izotov2025}.  

Flux sensitivity is also an important property to consider when measuring $f_{\mathrm{esc}}^{\mathrm{LyC}}$.  As a lower limit, the observed flux (accounting for IGM absorption) at 900 \AA\ is projected to be about $10^{-19}$ erg/s/cm$^2$/\AA\ or about 30 absolute magnitudes for $f_{\rm esc}^{\rm LyC} = 0.3\%$ in typical SF galaxies.  These numbers drop by about one dex out to $z = 1$ and by two dex at $z = 3$ \cite{McCandliss2017}.      

The following lines will be useful in the study of galactic winds.

\subsection{Lyman Alpha}

The H I 1216 \AA\ Lyman alpha (Ly$\alpha$) line is the best known proxy of LyC emission.  A double peaked emission feature is often observed in SF galaxies with the peak separation showing a negative correlation with $f_{\rm esc}^{\rm LyC}$ \cite{Flury2022b,Izotov2016a,Izotov2016b,Izotov2018b,Izotov2021,Izotov2022}.  Ly$\alpha$ acts as a direct and sensitive probe of the neutral gas and dust distribution in galaxies and galactic outflows \cite{Verhamme2015,Chung2019}.  The Ly$\alpha$ absorption feature has been linked to the high velocity profile of various metal lines \cite{Erb2023}.  Rather surprisingly, Ly$\alpha$ emission is frequently observed at high redshifts \cite{Saxena2024}, likely following the escape through ionized bubbles \cite{Mason2020}, keeping the door open for its use as a probe of $f_{\mathrm{esc}}^{\mathrm{LyC}}$ at high redshifts.

\subsection{Lyman Series}

The Lyman series—spanning from Ly$\beta$ at 1026 \AA, Ly$\gamma$ at 972.5 \AA, Ly$\delta$ at 949.7 \AA, and Ly$\epsilon$ at 937.8 \AA\ down to the Lyman limit at 911.8 \AA—provides another direct probe of neutral hydrogen. Unlike Ly$\alpha$, the higher-order Lyman lines typically lack a strong nebular emission component and instead appear in absorption or as P Cygni profiles, offering a valuable window into the physical state of the ISM and outflows. Their generally lower oscillator strengths at short wavelengths (e.g., Ly$\beta$–Ly$\epsilon$: 0.079–0.008) lead to lower optical depths, creating more favorable conditions for radiative transfer modeling. The absorption spectra of the Lyman series also present an ideal comparison to metal absorption lines, enabling studies of how metals trace neutral hydrogen. Although the high-energy end of the series often suffers from line blending and foreground contamination from the Milky Way at low redshifts, relatively clean Lyman series absorption features have been observed at redshifts $\sim$0.3 with COS \cite{Henry2015,Carr2025Sup}.

\subsection{LIS Metal Lines}

Low-ionization state (LIS) metal lines such as Si II 1190 \AA, 1193 \AA, 1265 \AA, 1393 \AA, 1527 \AA\ act as direct probes of the cold clouds ($10^4$ K) that trace the winds of galaxies and ISM.  Si II lines, in particular, have been explored as potential diagnostics of LyC escape \cite{Chisholm2017a}.  Moreover, when combined with the higher ionization states lines of Si III 1206 \AA\ and Si IV 1393 \AA, 1403 \AA, Si II lines have been used to map the ionization structure of galactic winds and constrain the abundances of neutral and total hydrogen \cite{Carr2021a,Xu2022a,Huberty2024}.  Additional lines include N II 1084 \AA, O I 1302 \AA, and C II 1334 \AA.  Like the Lyman series, these lines also appear in absorption or as P Cygni profiles.  As metal lines, they typically have much lower column densities than H I, creating ideal conditions for radiative transfer modeling.

Metal lines with fluorescent components, such as Si II 1190 \AA\ (Si II* 1194 \AA) and Si II 1193 \AA\ (Si II* 1197 \AA), offer an ideal opportunity to probe pure absorption profiles, absent blue emission infilling \cite{Prochaska2011,Scarlata2015}.  While resonant emission can fill in absorption wells, fluorescent emission will appear away at longer wavelengths.  Therefore, fluorescent lines offer an ideal opportunity to distinguish different absorption components, including the redshifted absorption signatures of galactic inflows \cite{Carr2022}.

\subsection{WIS Metal Lines}

Warm-ionization state (WIS) metal tracers such as N III 990 \AA, C III 977 \AA,  O VI 1032 \AA, 1038 \AA, Si IV 1394 \AA, 1403 \AA, C IV 1548 \AA, 1551 \AA\ offer a potential probe of the warm-hot ($10^{5.5}$ K) phase of galactic winds.  These lines offer a constraint on the most rapidly cooling gas in the CGM, that is prime for re-accretion onto the galaxy, making them ideal probes for how galactic inflows may affect LyC escape \cite{Marques-Chaves2022b}.  C IV, in particular, has been linked to catastrophic cooling and suppressed winds \cite{Oey2023}.  When combined with LIS lines, WIS lines can be used to further map the ionization structure of the winds, where the majority of mass is typically believed to occupy the cold + warm phases \cite{Kim2020a,Kim2020b,Li2020}.     

Galactic winds can extend tens to hundreds of kiloparsecs into the CGM, with the cool phase generally reaching shorter distances \cite{Carr2021a,Xu2022a}. For LzLCS galaxies, the mean Ly$\alpha$ half-light radius $r_{50} = 3.3\,\mathrm{kpc}$, while the mean 90\% light radius $r_{90} = 11.7\,\mathrm{kpc}$ \cite{Saldana-Lopez2025}. Observations with COS at low redshift are limited by its $1.25^{\prime\prime}$ aperture, which subtends a physical scale of only 3.4 kpc at $z = 0.22$ (e.g., \cite{James2022}).  

\subsection{FUV spectral energy distribution}

Modeling the FUV spectral energy distribution (SED) using BPASS or Starburst99 can provide valuable insights into the relative ages of stellar populations, which in turn can help identify the dominant source of feedback (i.e., radiation vs. supernovae; \cite{Flury2025,Carr2025LyC,Carr2025Sup}), metallicities, and star formation histories \cite{Chisholm2019}.  Key diagnostic features include the stellar wind lines OVI 1307 \AA\ and N V 1240 \AA, which are particularly sensitive to stellar population age \cite{Chisholm2022}, and the photospheric line C III 117 \AA, which is sensitive to metallicity \cite{Saldana-Lopez2022}.  The aforementioned lines can be well captured at low resolution $R = 1000$ \cite{Saldana-Lopez2022}, but higher resolution $\rm R = 3000-5000$ is required to distinguish absorption in galactic winds, particularly in photospheric lines \cite{Carr2025Sup}.  Photospheric lines generally fall between $1000-1800$ \AA\  and wind lines from $1030-1550$ \AA.

\section{Discussion and Requirements}

In this section, we discuss the prospects for future instrumentation enabled by the HWO and the observational requirements needed to resolve the origins and pathways of LyC escape in nearby SF galaxies. We delineate the science into three tiers—state-of-the-art, moderate progress, and transformational capability—with the latter representing a substantial advance beyond what is currently feasible with HST.

The COS spectrograph currently enables medium-resolution integrated spectroscopy of star-forming galaxies. For example, using the G130M grating, achieving a $\mathrm{S/N} = 10$ at a spectral resolution of $R = 4000$ or $\sim 75\ \rm km\ s^{-1}$ near 1000~\AA\ requires $\sim$ 11 hours (12 HST orbits) when observing the star-forming galaxy J115205+340050, a known LyC emitter ($f_{\mathrm{esc}}^{\mathrm{LyC}} = 13\%$, $z = 0.3419$; \cite{Flury2022a}). While feasible, such exposure times are costly, contributing to the scarcity of high-resolution, high-S/N spectra of LyC leakers in the literature. However, the few observations that meet these requirements reveal multiple kinematic components in galactic wind absorption lines, demonstrating that additional and physically meaningful structure appears when this resolution is reached \cite{Carr2025Sup}.

Tests against simulations show that the properties of cool galactic outflows–column density ($\sigma \sim 1.3$ dex), opening angle ($\sigma \sim 10$ deg), and flow rates ($\sigma \sim 0.5$ dex)–can be teased out from spectral lines at a resolution $\rm R = 15,000$ and $\rm S/N = 10$.  Similar studies have also proposed reaching $\rm R = 6000$ to capture the absorption properties of galactic inflows \cite{Carr2022}.  At higher resolution–$\rm R = 30,000 - 100,000$–we would be able to probe spectra down to the turbulent/thermal velocity regime.  This is the smallest scale at which additional structure could be detected, potentially revealing the cold clouds or multiphase structure of the CGM \cite{Fielding2022}.    

Using the HWO UV Spectrograph simulator\footnote{\url{https://hwo.stsci.edu/uvspec_etc}}, we estimate that a $\mathrm{S/N} = 10$ at a spectral resolution of $\sim 75\ \rm km\ s^{-1}$ near 1000~\AA\ can be achieved in just $\sim$ 24 (13) minutes with a 6 (8)-meter aperture.  At this level of gathering power, what is now state-of-the-art would become routine with HWO.  However, to enable truly transformational science and identify how LyC photons escape from galaxies, high-resolution spectral imaging is required. To image SSCs (1-100 pc) at redshift $z = 0.22$, a resolution of 0.28-28 mas is required.  These values become 0.36-36 mas at $z = 0.16$.  A 6 (8)-meter aperture becomes diffraction limited at 4.6 (3.5) mas at 1100 \AA, pushing us close to what is possible.  A $3^{\prime\prime} \times 3^{\prime\prime}$ ($\sim 10.6$ kpc at $z = 0.22$) field of view (FOV) would allow us to capture the majority of the Ly$\alpha$ halo. To capture the full CGM, an even larger field of view ($5-10^{\prime\prime}$) would be required. 

A UV IFS would be ideal, providing optimal conditions for radiative transfer modeling (e.g., \cite{Burchett2021,Erb2023}). A multi-object spectrograph (MOS) would also be valuable, requiring observations of $\sim$ 20–50 targets per galaxy to capture all LyC-emitting clusters. In practice, a combination of instruments will likely be required to fully characterize how LyC radiation escapes from local star-forming galaxies. An IFS with a spectral resolution of $\rm R \sim 5{,}000$, sub–30 mas spatial resolution, and a $3^{\prime\prime} \times 3^{\prime\prime}$ field of view would enable transformational advances. Higher spectral resolution ($\rm R \sim 30{,}000)$ may be more readily achieved with a MOS, while probing the highest-resolution kinematics of the CGM may require an échelle spectrograph. A summary of the observational requirements is provided in Table~\ref{tab:summary}. 

\begin{table*}[!ht]
\caption{Observational Capabilities and Requirements for Resolving LyC Escape}
\label{tab:summary}
\smallskip
\centering
\resizebox{\textwidth}{!}{%
\begin{tabular}{|p{3.5cm}|p{3.2cm}|p{3.2cm}|p{3.5cm}|p{4.2cm}|}
\hline
\textbf{Observation / Requirement} 
& \textbf{State-of-the-Art} 
& \textbf{Moderate Progress} 
& \textbf{Transformational Capability} 
& \textbf{Justification / Comments} \\
\hline

UV Spectral Coverage (Observed frame) 
& COS: $\sim$1100--1800~\AA 
& 990--2000~\AA 
& $<990$--2000~\AA 
& Moderate progress enables direct LyC detection at $z=0.16$, below this number allows for the detection of nebular contributions.  Higher wavelengths allow access to key metal lines (Si~II, Ly$\alpha$, C~IV) tracing multiphase outflows and CGM structure. \\
\hline

Spectral Resolution 
& $R \sim 4{,}000$  
& $R \sim 6{,}000$--15{,}000 
& $R \sim 30{,}000$--100{,}000 
& $R\gtrsim15{,}000$ resolves bulk outflow properties; $R\gtrsim30{,}000$ probes turbulent and thermal velocity scales of cool clouds and CGM substructure. \\
\hline

Signal-to-Noise 
& $\mathrm{S/N}\sim10$ in $\sim$11 hr at $\rm R = 4{,}000$ (HST/COS)
& $\mathrm{S/N}\sim10$ in minutes at $\rm R = 4{,}000$ (HWO, 6--8 m) 
& High $\mathrm{S/N}$ at high $R$ 
& High $\mathrm{S/N}$ is required to identify weak absorption components and subtle profile structure linked to cloud-scale physics. \\
\hline

Spatial Resolution 
& Integrated spectroscopy 
& $\sim$30--50 mas 
& $<30$ mas (diffraction-limited) 
& Resolving SSCs (1--100 pc) requires 0.3--30 mas at $z=0.22$; a 6 (8) m telescope reaches 4.6 (3.5) mas at 1100~\AA. \\
\hline

Field of View 
& COS aperture: $1.25^{\prime\prime}$ 
& $3^{\prime\prime} \times 3^{\prime\prime}$ 
& $5^{\prime\prime} \times 10^{\prime\prime}$ 
& $3^{\prime\prime}$ captures the Ly$\alpha$ halo ($\sim$10 kpc at $z=0.22$); larger FOV required to map the full CGM and extended winds. \\
\hline

Instrument Architecture 
& Single-aperture spectroscopy 
& UV IFS or MOS 
& UV IFS + MOS + échelle 
& IFS enables radiative transfer modeling and spatially resolved LyC escape; MOS supports high-$R$ surveys of SSCs; échelle required for highest resolution studies. \\
\hline

\end{tabular}
}
\end{table*}

\subsection*{Disclosures}
The authors have nothing to declare and report that they have no potential conflicts of interest.


\subsection* {Code, Data, and Materials Availability}

Data used to make Figure~\ref{fig:diagnostics} can be found in \cite{Flury2022b,Flury2022_erratum}.


\subsection* {Acknowledgments}
C.~C. is supported by NSFC grant W2433001 and the NSFC Talent-Introduction Program.  R.~C. acknowledges in part financial support from the start-up funding of Zhejiang University and Zhejiang provincial top level research support program.  OpenAI ChatGPT (version 4.0) was used to assist with language and phrasing of the manuscript.


\bibliography{report}   

@ARTICLE{Amorin2024,
       author = {{Amor{\'\i}n}, R.~O. and {Rodr{\'\i}guez-Henr{\'\i}quez}, M. and {Fern{\'a}ndez}, V. and {V{\'\i}lchez}, J.~M. and {Marques-Chaves}, R. and {Schaerer}, D. and {Izotov}, Y.~I. and {Firpo}, V. and {Guseva}, N. and {Jaskot}, A.~E. and {Komarova}, L. and {Mu{\~n}oz-Vergara}, D. and {Oey}, M.~S. and {Bait}, O. and {Carr}, C. and {Chisholm}, J. and {Ferguson}, H. and {Flury}, S.~R. and {Giavalisco}, M. and {Hayes}, M.~J. and {Henry}, A. and {Ji}, Z. and {King}, W. and {Leclercq}, F. and {{\"O}stlin}, G. and {Pentericci}, L. and {Saldana-Lopez}, A. and {Thuan}, T.~X. and {Trebitsch}, M. and {Wang}, B. and {Worseck}, G. and {Xu}, X.},
        title = "{Ubiquitous broad-line emission and the relation between ionized gas outflows and Lyman continuum escape in Green Pea galaxies}",
      journal = {Astronomy \& Astrophysics},
         year = 2024,
        month = feb,
       volume = {682},
          eid = {L25},
        pages = {L25},
          doi = {10.1051/0004-6361/202449175},
archivePrefix = {arXiv},
       eprint = {2401.04278},
 primaryClass = {astro-ph.GA},
       adsurl = {https://ui.adsabs.harvard.edu/abs/2024A&A...682L..25A}
}

@article{Atek2024,
  title={Most of the photons that reionized the Universe came from dwarf galaxies},
  author={Atek, Hakim and Labb{\'e}, Ivo and Furtak, Lukas J and Chemerynska, Iryna and Fujimoto, Seiji and Setton, David J and Miller, Tim B and Oesch, Pascal and Bezanson, Rachel and Price, Sedona H and others},
  journal={Nature},
  volume={626},
  number={8001},
  pages={975--978},
  year={2024},
  publisher={Nature Publishing Group UK London}
}

@ARTICLE{Bait2024,
       author = {{Bait}, Omkar and {Borthakur}, Sanchayeeta and {Schaerer}, Daniel and {Momjian}, Emmanuel and {Sebastian}, Biny and {Saldana-Lopez}, Alberto and {Flury}, Sophia R. and {Chisholm}, John and {Marques-Chaves}, Rui and {Jaskot}, Anne E. and {Ferguson}, Harry C. and {Worseck}, Gabor and {Ji}, Zhiyuan and {Komarova}, Lena and {Trebitsch}, Maxime and {Hayes}, Matthew J. and {Pentericci}, Laura and {Ostlin}, Goran and {Thuan}, Trinh and {Amor{\'\i}n}, Ricardo O. and {Wang}, Bingjie and {Xu}, Xinfeng and {Sargent}, Mark T.},
        title = "{Low-redshift Lyman Continuum Survey (LzLCS). Radio continuum properties of low-z Lyman continuum emitters}",
      journal = {Astronomy \& Astrophysics},
         year = 2024,
        month = aug,
       volume = {688},
          eid = {A198},
        pages = {A198},
          doi = {10.1051/0004-6361/202348416},
archivePrefix = {arXiv},
       eprint = {2310.18817},
 primaryClass = {astro-ph.GA},
       adsurl = {https://ui.adsabs.harvard.edu/abs/2024A&A...688A.198B}
}

@article{Borthakur2014,
  title={A local clue to the reionization of the universe},
  author={Borthakur, Sanchayeeta and Heckman, Timothy M and Leitherer, Claus and Overzier, Roderik A},
  journal={Science},
  volume={346},
  number={6206},
  pages={216--219},
  year={2014},
  publisher={American Association for the Advancement of Science}
}

@ARTICLE{James2022,
       author = {{James}, Bethan L. and {Berg}, Danielle A. and {King}, Teagan and {Sahnow}, David J. and {Mingozzi}, Matilde and {Chisholm}, John and {Heckman}, Timothy and {Martin}, Crystal L. and {Stark}, Dan P. and {Aloisi}, Alessandra and {Amor{\'\i}n}, Ricardo O. and {Arellano-C{\'o}rdova}, Karla Z. and {Bayliss}, Matthew and {Bordoloi}, Rongmon and {Brinchmann}, Jarle and {Charlot}, St{\'e}phane and {Chen}, Zuyi and {Chevallard}, Jacopo and {Clark}, Ilyse and {Erb}, Dawn K. and {Feltre}, Anna and {Hayes}, Matthew and {Henry}, Alaina and {Hernandez}, Svea and {Jaskot}, Anne and {Kewley}, Lisa J. and {Kumari}, Nimisha and {Leitherer}, Claus and {Llerena}, Mario and {Maseda}, Michael and {Nanayakkara}, Themiya and {Ouchi}, Masami and {Plat}, Adele and {Pogge}, Richard W. and {Ravindranath}, Swara and {Rigby}, Jane R. and {Scarlata}, Claudia and {Senchyna}, Peter and {Skillman}, Evan D. and {Steidel}, Charles C. and {Strom}, Allison L. and {Sugahara}, Yuma and {Wilkins}, Stephen M. and {Wofford}, Aida and {Xu}, Xinfeng and {Classy Team}},
        title = "{CLASSY. II. A Technical Overview of the COS Legacy Archive Spectroscopic Survey}",
      journal = {The Astrophysical Journal Supplement Series},
         year = 2022,
        month = oct,
       volume = {262},
       number = {2},
          eid = {37},
        pages = {37},
          doi = {10.3847/1538-4365/ac8008},
archivePrefix = {arXiv},
       eprint = {2206.01224},
 primaryClass = {astro-ph.GA},
       adsurl = {https://ui.adsabs.harvard.edu/abs/2022ApJS..262...37J}
}

@ARTICLE{Burchett2021,
       author = {{Burchett}, Joseph N. and {Rubin}, Kate H.~R. and {Prochaska}, J. Xavier and {Coil}, Alison L. and {Vaught}, Ryan Rickards and {Hennawi}, Joseph F.},
        title = "{Circumgalactic Mg II Emission from an Isotropic Starburst Galaxy Outflow Mapped by KCWI}",
      journal = {The Astrophysical Journal},
         year = 2021,
        month = mar,
       volume = {909},
       number = {2},
          eid = {151},
        pages = {151},
          doi = {10.3847/1538-4357/abd4e0},
archivePrefix = {arXiv},
       eprint = {2005.03017},
 primaryClass = {astro-ph.GA},
       adsurl = {https://ui.adsabs.harvard.edu/abs/2021ApJ...909..151B}
}

@ARTICLE{Carr2018,
       author = {{Carr}, Cody and {Scarlata}, Claudia and {Panagia}, Nino and {Henry}, Alaina},
        title = "{A Semi-analytical Line Transfer (SALT) Model. II: The Effects of a Bi-conical Geometry}",
      journal = {The Astrophysical Journal},
         year = 2018,
        month = jun,
       volume = {860},
       number = {2},
          eid = {143},
        pages = {143},
          doi = {10.3847/1538-4357/aac48e},
archivePrefix = {arXiv},
       eprint = {1805.05981},
 primaryClass = {astro-ph.GA},
       adsurl = {https://ui.adsabs.harvard.edu/abs/2018ApJ...860..143C}
}

@ARTICLE{Carr2021a,
       author = {{Carr}, Cody and {Scarlata}, Claudia and {Henry}, Alaina and {Panagia}, Nino},
        title = "{The Effects of Biconical Outflows on Ly{\ensuremath{\alpha}} Escape from Green Peas}",
      journal = {The Astrophysical Journal},
         year = 2021,
        month = jan,
       volume = {906},
       number = {2},
          eid = {104},
        pages = {104},
          doi = {10.3847/1538-4357/abc7c3},
archivePrefix = {arXiv},
       eprint = {2011.02549},
 primaryClass = {astro-ph.GA},
       adsurl = {https://ui.adsabs.harvard.edu/abs/2021ApJ...906..104C}
}

@ARTICLE{Carr2022,
       author = {{Carr}, C. and {Scarlata}, C.},
        title = "{A Semianalytical Line Transfer Model. III. Galactic Inflows}",
      journal = {The Astrophysical Journal},
         year = 2022,
        month = nov,
       volume = {939},
       number = {1},
          eid = {47},
        pages = {47},
          doi = {10.3847/1538-4357/ac93fa},
archivePrefix = {arXiv},
       eprint = {2209.14485},
 primaryClass = {astro-ph.GA},
       adsurl = {https://ui.adsabs.harvard.edu/abs/2022ApJ...939...47C}
}

@ARTICLE{Carr2023,
       author = {{Carr}, C. and {Michel-Dansac}, L. and {Blaizot}, J. and {Scarlata}, C. and {Henry}, A. and {Verhamme}, A.},
        title = "{Testing SALT Approximations with Numerical Radiation Transfer Code. I. Validity and Applicability}",
      journal = {The Astrophysical Journal},
         year = 2023,
        month = jul,
       volume = {952},
       number = {1},
          eid = {88},
        pages = {88},
          doi = {10.3847/1538-4357/acd331},
archivePrefix = {arXiv},
       eprint = {2209.14473},
 primaryClass = {astro-ph.GA},
       adsurl = {https://ui.adsabs.harvard.edu/abs/2023ApJ...952...88C}
}

@article{Carr2025LyC,
doi = {10.3847/1538-4357/adb72f},
url = {https://dx.doi.org/10.3847/1538-4357/adb72f},
year = {2025},
month = {mar},
publisher = {The American Astronomical Society},
volume = {982},
number = {2},
pages = {137},
author = {Carr, Cody A. and Cen, Renyue and Scarlata, Claudia and Xu, Xinfeng and Henry, Alaina and Marques-Chaves, Rui and Schaerer, Daniel and Amorín, Ricardo O. and Oey, M. S. and Komarova, Lena and Flury, Sophia and Jaskot, Anne and Saldana-Lopez, Alberto and Ji, Zhiyuan and Huberty, Mason and Heckman, Timothy and Östlin, Göran and Bait, Omkar and Hayes, Matthew James and Thuan, Trinh and Ravindranath, Swara and Berg, Danielle A. and Giavalisco, Mauro and Rutkowski, Michael and Borthakur, Sanchayeeta and Chisholm, John and Ferguson, Harry C. and Michel-Dansac, Leo and Verhamme, Anne and Worseck, Gábor},
title = {The Effect of Radiation and Supernovae Feedback on LyC Escape in Local Star-forming Galaxies},
journal = {The Astrophysical Journal}
}

@ARTICLE{Carr2025MOR,
       author = {{Carr}, Cody A. and {Smith}, Aaron and {Pandya}, Viraj and {Hayward}, Christopher C. and {Huberty}, Mason and {Scarlata}, Claudia and {Cen}, Renyue},
        title = "{Evaluating Mass Outflow Rate Estimators in FIRE-2 Simulations: Toward a Self-consistent Framework for Spectral-line-based Predictions}",
      journal = {The Astrophysical Journal},
         year = 2025,
        month = sep,
       volume = {990},
       number = {2},
          eid = {220},
        pages = {220},
          doi = {10.3847/1538-4357/adf4c2},
archivePrefix = {arXiv},
       eprint = {2503.22312},
 primaryClass = {astro-ph.GA},
       adsurl = {https://ui.adsabs.harvard.edu/abs/2025ApJ...990..220C}
}

@ARTICLE{Carr2025Sup,
       author = {{Carr}, Cody and {Cen}, Renyue and {McCandliss}, Stephan and {Ford}, Jack and {Saldana-Lopez}, Alberto and {Scarlata}, Claudia and {Huberty}, Mason and {Jaskot}, Anne and {Flury}, Sophia and {Oey}, M.~S. and {Amor{\'\i}n}, Ricardo O. and {Borthakur}, Sanchayeeta and {Hayes}, Matthew and {Heckman}, Timothy and {Ji}, Zhiyuan and {Komarova}, Lena and {Le Reste}, Alexandra and {Leclercq}, Floriane and {Marques-Chaves}, Rui and {Michel-Dansac}, Leo and {{\"O}stlin}, G{\"o}ran and {Ravindranath}, Swara and {Rutkowski}, Michael J. and {Schaerer}, Daniel and {Thuan}, Trinh and {Vanzella}, Eros and {Wang}, Bingjie and {Xu}, Xinfeng},
        title = "{Supernovae Driven Winds Impede Lyman Continuum Escape from Dwarf Galaxies in First 10 Myr}",
      journal = {arXiv e-prints},
         year = 2025,
        month = oct,
          eid = {arXiv:2510.21197},
        pages = {arXiv:2510.21197},
          doi = {10.48550/arXiv.2510.21197},
archivePrefix = {arXiv},
       eprint = {2510.21197},
 primaryClass = {astro-ph.GA},
       adsurl = {https://ui.adsabs.harvard.edu/abs/2025arXiv251021197C}
}

@ARTICLE{Cen2020,
       author = {{Cen}, Renyue},
        title = "{Physics of Prodigious Lyman Continuum Leakers}",
      journal = {The Astrophysical Journal Letters},
         year = 2020,
        month = jan,
       volume = {889},
       number = {1},
          eid = {L22},
        pages = {L22},
          doi = {10.3847/2041-8213/ab6560},
archivePrefix = {arXiv},
       eprint = {2001.11083},
 primaryClass = {astro-ph.GA},
       adsurl = {https://ui.adsabs.harvard.edu/abs/2020ApJ...889L..22C}
}

@ARTICLE{Chisholm2016b,
       author = {{Chisholm}, John and {Tremonti Christy}, A. and {Leitherer}, Claus and {Chen}, Yanmei},
        title = "{A robust measurement of the mass outflow rate of the galactic outflow from NGC 6090}",
      journal = {Monthly Notices of the Royal Astronomical Society},
         year = 2016,
        month = nov,
       volume = {463},
       number = {1},
        pages = {541-556},
          doi = {10.1093/mnras/stw1951},
archivePrefix = {arXiv},
       eprint = {1605.05769},
 primaryClass = {astro-ph.GA},
       adsurl = {https://ui.adsabs.harvard.edu/abs/2016MNRAS.463..541C}
}

@article{Chisholm2017a,
  title={Do galaxies that leak ionizing photons have extreme outflows?},
  author={Chisholm, John and Orlitov{\'a}, I and Schaerer, D and Verhamme, A and Worseck, G and Izotov, YI and Thuan, TX and Guseva, NG},
  journal={Astronomy \& Astrophysics},
  volume={605},
  pages={A67},
  year={2017},
  publisher={EDP Sciences}
}

@article{Chisholm2019,
  title={Constraining the metallicities, ages, star formation histories, and ionizing continua of extragalactic massive star populations*},
  author={Chisholm, J and Rigby, JR and Bayliss, M and Berg, DA and Dahle, H{\aa}kon and Gladders, M and Sharon, K},
  journal={The Astrophysical Journal},
  volume={882},
  number={2},
  pages={182},
  year={2019},
  publisher={IOP Publishing}
}

@ARTICLE{Chisholm2022,
       author = {{Chisholm}, J. and {Saldana-Lopez}, A. and {Flury}, S. and {Schaerer}, D. and {Jaskot}, A. and {Amor{\'\i}n}, R. and {Atek}, H. and {Finkelstein}, S.~L. and {Fleming}, B. and {Ferguson}, H. and {Fern{\'a}ndez}, V. and {Giavalisco}, M. and {Hayes}, M. and {Heckman}, T. and {Henry}, A. and {Ji}, Z. and {Marques-Chaves}, R. and {Mauerhofer}, V. and {McCandliss}, S. and {Oey}, M.~S. and {{\"O}stlin}, G. and {Rutkowski}, M. and {Scarlata}, C. and {Thuan}, T. and {Trebitsch}, M. and {Wang}, B. and {Worseck}, G. and {Xu}, X.},
        title = "{The far-ultraviolet continuum slope as a Lyman Continuum escape estimator at high redshift}",
      journal = {Monthly Notices of the Royal Astronomical Society},
         year = 2022,
        month = dec,
       volume = {517},
       number = {4},
        pages = {5104-5120},
          doi = {10.1093/mnras/stac2874},
archivePrefix = {arXiv},
       eprint = {2207.05771},
 primaryClass = {astro-ph.GA},
       adsurl = {https://ui.adsabs.harvard.edu/abs/2022MNRAS.517.5104C}
}

@ARTICLE{Choustikov2024,
       author = {{Choustikov}, Nicholas and {Katz}, Harley and {Saxena}, Aayush and {Cameron}, Alex J. and {Devriendt}, Julien and {Slyz}, Adrianne and {Rosdahl}, Joki and {Blaizot}, Jeremy and {Michel-Dansac}, Leo},
        title = "{The Physics of Indirect Estimators of Lyman Continuum Escape and their Application to High-Redshift JWST Galaxies}",
      journal = {Monthly Notices of the Royal Astronomical Society},
         year = 2024,
        month = apr,
       volume = {529},
       number = {4},
        pages = {3751-3767},
          doi = {10.1093/mnras/stae776},
archivePrefix = {arXiv},
       eprint = {2304.08526},
 primaryClass = {astro-ph.GA},
       adsurl = {https://ui.adsabs.harvard.edu/abs/2024MNRAS.529.3751C}
}

@ARTICLE{Chung2019,
       author = {{Chung}, Andrew S. and {Dijkstra}, Mark and {Ciardi}, Benedetta and {Kakiichi}, Koki and {Naab}, Thorsten},
        title = "{The circumgalactic medium in Lyman {\ensuremath{\alpha}}: a new constraint on galactic outflow models}",
      journal = {Monthly Notices of the Royal Astronomical Society},
         year = 2019,
        month = apr,
       volume = {484},
       number = {2},
        pages = {2420-2432},
          doi = {10.1093/mnras/stz149},
archivePrefix = {arXiv},
       eprint = {1901.04015},
 primaryClass = {astro-ph.GA},
       adsurl = {https://ui.adsabs.harvard.edu/abs/2019MNRAS.484.2420C}
}

@ARTICLE{Eldridge2017,
       author = {{Eldridge}, J.~J. and {Stanway}, E.~R. and {Xiao}, L. and {McClelland}, L.~A.~S. and {Taylor}, G. and {Ng}, M. and {Greis}, S.~M.~L. and {Bray}, J.~C.},
        title = "{Binary Population and Spectral Synthesis Version 2.1: Construction, Observational Verification, and New Results}",
      journal = {Publications of the Astronomical Society of Australia},
         year = 2017,
        month = nov,
       volume = {34},
          eid = {e058},
        pages = {e058},
          doi = {10.1017/pasa.2017.51},
archivePrefix = {arXiv},
       eprint = {1710.02154},
 primaryClass = {astro-ph.SR},
       adsurl = {https://ui.adsabs.harvard.edu/abs/2017PASA...34...58E}
}

@ARTICLE{Erb2023,
       author = {{Erb}, Dawn K. and {Li}, Zhihui and {Steidel}, Charles C. and {Chen}, Yuguang and {Gronke}, Max and {Strom}, Allison L. and {Trainor}, Ryan F. and {Rudie}, Gwen C.},
        title = "{The Circumgalactic Medium of Extreme Emission Line Galaxies at z 2: Resolved Spectroscopy and Radiative Transfer Modeling of Spatially Extended Ly{\ensuremath{\alpha}} Emission in the KBSS-KCWI Survey}",
      journal = {The Astrophysical Journal},
         year = 2023,
        month = aug,
       volume = {953},
       number = {1},
          eid = {118},
        pages = {118},
          doi = {10.3847/1538-4357/acd849},
archivePrefix = {arXiv},
       eprint = {2210.02465},
 primaryClass = {astro-ph.GA},
       adsurl = {https://ui.adsabs.harvard.edu/abs/2023ApJ...953..118E}
}

@ARTICLE{Fielding2022,
       author = {{Fielding}, Drummond B. and {Bryan}, Greg L.},
        title = "{The Structure of Multiphase Galactic Winds}",
      journal = {The Astrophysical Journal},
         year = 2022,
        month = jan,
       volume = {924},
       number = {2},
          eid = {82},
        pages = {82},
          doi = {10.3847/1538-4357/ac2f41},
archivePrefix = {arXiv},
       eprint = {2108.05355},
 primaryClass = {astro-ph.GA},
       adsurl = {https://ui.adsabs.harvard.edu/abs/2022ApJ...924...82F}
}

@ARTICLE{Flury2022a,
       author = {{Flury}, Sophia R. and {Jaskot}, Anne E. and {Ferguson}, Harry C. and {Worseck}, G{\'a}bor and {Makan}, Kirill and {Chisholm}, John and {Saldana-Lopez}, Alberto and {Schaerer}, Daniel and {McCandliss}, Stephan and {Wang}, Bingjie and {Ford}, N.~M. and {Heckman}, Timothy and {Ji}, Zhiyuan and {Giavalisco}, Mauro and {Amorin}, Ricardo and {Atek}, Hakim and {Blaizot}, Jeremy and {Borthakur}, Sanchayeeta and {Carr}, Cody and {Castellano}, Marco and {Cristiani}, Stefano and {De Barros}, Stephane and {Dickinson}, Mark and {Finkelstein}, Steven L. and {Fleming}, Brian and {Fontanot}, Fabio and {Garel}, Thibault and {Grazian}, Andrea and {Hayes}, Matthew and {Henry}, Alaina and {Mauerhofer}, Valentin and {Micheva}, Genoveva and {Oey}, M.~S. and {Ostlin}, Goran and {Papovich}, Casey and {Pentericci}, Laura and {Ravindranath}, Swara and {Rosdahl}, Joakim and {Rutkowski}, Michael and {Santini}, Paola and {Scarlata}, Claudia and {Teplitz}, Harry and {Thuan}, Trinh and {Trebitsch}, Maxime and {Vanzella}, Eros and {Verhamme}, Anne and {Xu}, Xinfeng},
        title = "{The Low-redshift Lyman Continuum Survey. I. New, Diverse Local Lyman Continuum Emitters}",
      journal = {The Astrophysical Journal Supplement Series},
         year = 2022,
        month = may,
       volume = {260},
       number = {1},
          eid = {1},
        pages = {1},
          doi = {10.3847/1538-4365/ac5331},
archivePrefix = {arXiv},
       eprint = {2201.11716},
 primaryClass = {astro-ph.GA},
       adsurl = {https://ui.adsabs.harvard.edu/abs/2022ApJS..260....1F}
}

@ARTICLE{Flury2022_erratum,
       author = {{Flury}, Sophia R. and {Jaskot}, Anne E. and {Ferguson}, Harry C. and {Worseck}, Gabor and {Makan}, Kirill and {Chisholm}, John and {Saldana-Lopez}, Alberto and {Schaerer}, Daniel and {McCandliss}, Stephan R. and {Wang}, Bingjie and {Ford}, N.~M. and {Heckman}, Timothy and {Ji}, Zhiyuan and {Giavalisco}, Mauro and {Amor{\'\i}n}, Ricardo and {Atek}, Hakim and {Blaizot}, Jeremy and {Borthakur}, Sanchayeeta and {Carr}, Cody and {Castellano}, Marco and {Cristiani}, Stefano and {De Barros}, Stephane and {Dickinson}, Mark and {Finkelstein}, Steven L. and {Fleming}, Brian and {Fontanot}, Fabio and {Garel}, Thibault and {Grazian}, Andrea and {Hayes}, Matthew and {Henry}, Alaina and {Mauerhofer}, Valentin and {Micheva}, Genoveva and {Oey}, M.~S. and {Ostlin}, Goran and {Papovich}, Casey and {Pentericci}, Laura and {Ravindranath}, Swara and {Rosdahl}, Joakim and {Rutkowski}, Michael and {Santini}, Paola and {Scarlata}, Claudia and {Teplitz}, Harry and {Thuan}, Trinh and {Trebitsch}, Maxime and {Vanzella}, Eros and {Verhamme}, Anne and {Xu}, Xinfeng},
        title = "{Erratum: ``The Low-redshift Lyman Continuum Survey. I. New, Diverse Local Lyman Continuum Emitters'' (2022, ApJS, 260, 1)}",
      journal = {The Astrophysical Journal Supplement Series},
         year = 2024,
        month = dec,
       volume = {275},
       number = {2},
          eid = {47},
        pages = {47},
          doi = {10.3847/1538-4365/ad7bb9},
       adsurl = {https://ui.adsabs.harvard.edu/abs/2024ApJS..275...47F}
}

@ARTICLE{Flury2022b,
       author = {{Flury}, Sophia R. and {Jaskot}, Anne E. and {Ferguson}, Harry C. and {Worseck}, G{\'a}bor and {Makan}, Kirill and {Chisholm}, John and {Saldana-Lopez}, Alberto and {Schaerer}, Daniel and {McCandliss}, Stephan R. and {Xu}, Xinfeng and {Wang}, Bingjie and {Oey}, M.~S. and {Ford}, N.~M. and {Heckman}, Timothy and {Ji}, Zhiyuan and {Giavalisco}, Mauro and {Amor{\'\i}n}, Ricardo and {Atek}, Hakim and {Blaizot}, Jeremy and {Borthakur}, Sanchayeeta and {Carr}, Cody and {Castellano}, Marco and {De Barros}, Stephane and {Dickinson}, Mark and {Finkelstein}, Steven L. and {Fleming}, Brian and {Fontanot}, Fabio and {Garel}, Thibault and {Grazian}, Andrea and {Hayes}, Matthew and {Henry}, Alaina and {Mauerhofer}, Valentin and {Micheva}, Genoveva and {Ostlin}, Goran and {Papovich}, Casey and {Pentericci}, Laura and {Ravindranath}, Swara and {Rosdahl}, Joakim and {Rutkowski}, Michael and {Santini}, Paola and {Scarlata}, Claudia and {Teplitz}, Harry and {Thuan}, Trinh and {Trebitsch}, Maxime and {Vanzella}, Eros and {Verhamme}, Anne},
        title = "{The Low-redshift Lyman Continuum Survey. II. New Insights into LyC Diagnostics}",
      journal = {The Astrophysical Journal},
         year = 2022,
        month = may,
       volume = {930},
       number = {2},
          eid = {126},
        pages = {126},
          doi = {10.3847/1538-4357/ac61e4},
archivePrefix = {arXiv},
       eprint = {2203.15649},
 primaryClass = {astro-ph.GA},
       adsurl = {https://ui.adsabs.harvard.edu/abs/2022ApJ...930..126F}
}

@article{Flury2023,
  title={Galactic outflow emission line profiles: evidence for dusty, radiatively driven ionized winds in Mrk 462},
  author={Flury, Sophia R and Moran, Edward C and Eleazer, Miriam},
  journal={Monthly Notices of the Royal Astronomical Society},
  volume={525},
  number={3},
  pages={4231--4242},
  year={2023},
  publisher={Oxford University Press}
}

@ARTICLE{Flury2025,
       author = {{Flury}, Sophia R. and {Jaskot}, Anne E. and {Saldana-Lopez}, Alberto and {Oey}, M.~S. and {Chisholm}, John and {Amor{\'\i}n}, Ricardo and {Bait}, Omkar and {Borthakur}, Sanchayeeta and {Carr}, Cody and {Ferguson}, Henry C. and {Giavalisco}, Mauro and {Hayes}, Matthew and {Heckman}, Timothy and {Henry}, Alaina and {Ji}, Zhiyuan and {Komarova}, Lena and {Leclercq}, Florian and {Le Reste}, Alexandra and {McCandliss}, Stephan and {Marques-Chaves}, Rui and {{\"O}stlin}, G{\"o}ran and {Pentericci}, Laura and {Ravindranath}, Swara and {Rutkowski}, Michael and {Scarlata}, Claudia and {Schaerer}, Daniel and {Thuan}, Trinh and {Trebitsch}, Maxime and {Vanzella}, Eros and {Verhamme}, Anne and {Wang}, Bingjie and {Worseck}, G{\'a}bor and {Xu}, Xinfeng},
        title = "{The Low-redshift Lyman Continuum Survey: The Roles of Stellar Feedback and Interstellar Medium Geometry in LyC Escape}",
      journal = {The Astrophysical Journal},
         year = 2025,
        month = may,
       volume = {985},
       number = {1},
          eid = {128},
        pages = {128},
          doi = {10.3847/1538-4357/adc305},
archivePrefix = {arXiv},
       eprint = {2409.12118},
 primaryClass = {astro-ph.GA},
       adsurl = {https://ui.adsabs.harvard.edu/abs/2025ApJ...985..128F}
}

@ARTICLE{Green2012,
       author = {{Green}, James C. and {Froning}, Cynthia S. and {Osterman}, Steve and {Ebbets}, Dennis and {Heap}, Sara H. and {Leitherer}, Claus and {Linsky}, Jeffrey L. and {Savage}, Blair D. and {Sembach}, Kenneth and {Shull}, J. Michael and {Siegmund}, Oswald H.~W. and {Snow}, Theodore P. and {Spencer}, John and {Stern}, S. Alan and {Stocke}, John and {Welsh}, Barry and {B{\'e}land}, St{\'e}phane and {Burgh}, Eric B. and {Danforth}, Charles and {France}, Kevin and {Keeney}, Brian and {McPhate}, Jason and {Penton}, Steven V. and {Andrews}, John and {Brownsberger}, Kenneth and {Morse}, Jon and {Wilkinson}, Erik},
        title = "{The Cosmic Origins Spectrograph}",
      journal = {The Astrophysical Journal},
         year = 2012,
        month = jan,
       volume = {744},
       number = {1},
          eid = {60},
        pages = {60},
          doi = {10.1088/0004-637X/744/1/6010.1086/141956},
archivePrefix = {arXiv},
       eprint = {1110.0462},
 primaryClass = {astro-ph.IM},
       adsurl = {https://ui.adsabs.harvard.edu/abs/2012ApJ...744...60G}
}

@ARTICLE{Hayes2023b,
       author = {{Hayes}, Matthew J. and {Runnholm}, Axel and {Scarlata}, Claudia and {Gronke}, Max and {Rivera-Thorsen}, T. Emil},
        title = "{Spectral shapes of the Ly {\ensuremath{\alpha}} emission from galaxies - II. The influence of stellar properties and nebular conditions on the emergent Ly {\ensuremath{\alpha}} profiles}",
      journal = {Monthly Notices of the Royal Astronomical Society},
         year = 2023,
        month = apr,
       volume = {520},
       number = {4},
        pages = {5903-5927},
          doi = {10.1093/mnras/stad477},
archivePrefix = {arXiv},
       eprint = {2302.04875},
 primaryClass = {astro-ph.GA},
       adsurl = {https://ui.adsabs.harvard.edu/abs/2023MNRAS.520.5903H}
}

@ARTICLE{Henry2015,
       author = {{Henry}, Alaina and {Scarlata}, Claudia and {Martin}, Crystal L. and {Erb}, Dawn},
        title = "{Ly{\ensuremath{\alpha}} Emission from Green Peas: The Role of Circumgalactic Gas Density, Covering, and Kinematics}",
      journal = {The Astrophysical Journal},
         year = 2015,
        month = aug,
       volume = {809},
       number = {1},
          eid = {19},
        pages = {19},
          doi = {10.1088/0004-637X/809/1/19},
archivePrefix = {arXiv},
       eprint = {1505.05149},
 primaryClass = {astro-ph.GA},
       adsurl = {https://ui.adsabs.harvard.edu/abs/2015ApJ...809...19H}
}

@article{Huberty2024,
  title={CLASSY. X. Highlighting Differences between Partial Covering and Semianalytic Modeling in the Estimation of Galactic Outflow Properties},
  author={Huberty, Mason and Carr, Cody and Scarlata, Claudia and Heckman, Timothy and Henry, Alaina and Xu, Xinfeng and Arellano-C{\'o}rdova, Karla Z and Berg, Danielle A and Charlot, St{\'e}phane and Chisholm, John and others},
  journal={The Astrophysical Journal},
  volume={975},
  number={1},
  pages={58},
  year={2024},
  publisher={IOP Publishing}
}

@ARTICLE{Inoue2010,
       author = {{Inoue}, Akio K.},
        title = "{Lyman `bump' galaxies - I. Spectral energy distribution of galaxies with an escape of nebular Lyman continuum}",
      journal = {Monthly Notices of the Royal Astronomical Society},
         year = 2010,
        month = jan,
       volume = {401},
       number = {2},
        pages = {1325-1333},
          doi = {10.1111/j.1365-2966.2009.15730.x},
archivePrefix = {arXiv},
       eprint = {0908.3925},
 primaryClass = {astro-ph.CO},
       adsurl = {https://ui.adsabs.harvard.edu/abs/2010MNRAS.401.1325I}
}

@ARTICLE{Inoue2014,
       author = {{Inoue}, Akio K. and {Shimizu}, Ikkoh and {Iwata}, Ikuru and {Tanaka}, Masayuki},
        title = "{An updated analytic model for attenuation by the intergalactic medium}",
      journal = {Monthly Notices of the Royal Astronomical Society},
         year = 2014,
        month = aug,
       volume = {442},
       number = {2},
        pages = {1805-1820},
          doi = {10.1093/mnras/stu936},
archivePrefix = {arXiv},
       eprint = {1402.0677},
 primaryClass = {astro-ph.CO},
       adsurl = {https://ui.adsabs.harvard.edu/abs/2014MNRAS.442.1805I}
}

@ARTICLE{Izotov2016a,
       author = {{Izotov}, Y.~I. and {Orlitov{\'a}}, I. and {Schaerer}, D. and {Thuan}, T.~X. and {Verhamme}, A. and {Guseva}, N.~G. and {Worseck}, G.},
        title = "{Eight per cent leakage of Lyman continuum photons from a compact, star-forming dwarf galaxy}",
      journal = {Nature},
         year = 2016,
        month = jan,
       volume = {529},
       number = {7585},
        pages = {178-180},
          doi = {10.1038/nature16456},
archivePrefix = {arXiv},
       eprint = {1601.03068},
 primaryClass = {astro-ph.GA},
       adsurl = {https://ui.adsabs.harvard.edu/abs/2016Natur.529..178I}
}

@ARTICLE{Izotov2016b,
       author = {{Izotov}, Y.~I. and {Schaerer}, D. and {Thuan}, T.~X. and {Worseck}, G. and {Guseva}, N.~G. and {Orlitov{\'a}}, I. and {Verhamme}, A.},
        title = "{Detection of high Lyman continuum leakage from four low-redshift compact star-forming galaxies}",
      journal = {Monthly Notices of the Royal Astronomical Society},
         year = 2016,
        month = oct,
       volume = {461},
       number = {4},
        pages = {3683-3701},
          doi = {10.1093/mnras/stw1205},
archivePrefix = {arXiv},
       eprint = {1605.05160},
 primaryClass = {astro-ph.GA},
       adsurl = {https://ui.adsabs.harvard.edu/abs/2016MNRAS.461.3683I}
}

@ARTICLE{Izotov2017,
       author = {{Izotov}, Y.~I. and {Thuan}, T.~X. and {Guseva}, N.~G.},
        title = "{LBT observations of compact star-forming galaxies with extremely high [O III]/[O II] flux ratios: He I emission-line ratios as diagnostics of Lyman continuum leakage}",
      journal = {Monthly Notices of the Royal Astronomical Society},
         year = 2017,
        month = oct,
       volume = {471},
       number = {1},
        pages = {548-561},
          doi = {10.1093/mnras/stx1629},
archivePrefix = {arXiv},
       eprint = {1706.08769},
 primaryClass = {astro-ph.GA},
       adsurl = {https://ui.adsabs.harvard.edu/abs/2017MNRAS.471..548I}
}

@ARTICLE{Izotov2018b,
       author = {{Izotov}, Y.~I. and {Worseck}, G. and {Schaerer}, D. and {Guseva}, N.~G. and {Thuan}, T.~X. and {Fricke}, Verhamme, A. and {Orlitov{\'a}}, I.},
        title = "{Low-redshift Lyman continuum leaking galaxies with high [O III]/[O II] ratios}",
      journal = {Monthly Notices of the Royal Astronomical Society},
         year = 2018,
        month = aug,
       volume = {478},
       number = {4},
        pages = {4851-4865},
          doi = {10.1093/mnras/sty1378},
archivePrefix = {arXiv},
       eprint = {1805.09865},
 primaryClass = {astro-ph.GA},
       adsurl = {https://ui.adsabs.harvard.edu/abs/2018MNRAS.478.4851I}
}

@ARTICLE{Izotov2020,
       author = {{Izotov}, Y.~I. and {Schaerer}, D. and {Worseck}, G. and {Verhamme}, A. and {Guseva}, N.~G. and {Thuan}, T.~X. and {Orlitov{\'a}}, I. and {Fricke}, K.~J.},
        title = "{Diverse properties of Ly {\ensuremath{\alpha}} emission in low-redshift compact star-forming galaxies with extremely high [O III]/[O II] ratios}",
      journal = {Monthly Notices of the Royal Astronomical Society},
         year = 2020,
        month = jan,
       volume = {491},
       number = {1},
        pages = {468-482},
          doi = {10.1093/mnras/stz3041},
archivePrefix = {arXiv},
       eprint = {1910.12773},
 primaryClass = {astro-ph.GA},
       adsurl = {https://ui.adsabs.harvard.edu/abs/2020MNRAS.491..468I}
}

@ARTICLE{Izotov2022,
       author = {{Izotov}, Y.~I. and {Chisholm}, J. and {Worseck}, G. and {Guseva}, N.~G. and {Schaerer}, D. and {Prochaska}, J.~X.},
        title = "{Lyman alpha and Lyman continuum emission of Mg II-selected star-forming galaxies}",
      journal = {Monthly Notices of the Royal Astronomical Society},
         year = 2022,
        month = sep,
       volume = {515},
       number = {2},
        pages = {2864-2881},
          doi = {10.1093/mnras/stac1899},
archivePrefix = {arXiv},
       eprint = {2207.04483},
 primaryClass = {astro-ph.GA},
       adsurl = {https://ui.adsabs.harvard.edu/abs/2022MNRAS.515.2864I}
}
\bibliographystyle{spiejour}   


\vspace{2ex}\noindent\textbf{Cody Carr} is currently participating in the Talent Introduction Program at Zhejiang University and will join the University of Michigan as the Extremely Large Telescope (ELT) Postdoctoral Fellow in 2026.  He received his PhD in astronomy from the University of Minnesota in 2023.  He has published multiple papers on topics related to galaxy formation, radiative transfer theory, UV spectroscopy, and reionization. 

\vspace{2ex}\noindent\textbf{Renyue Cen} is the director of the Center for Cosmology and Computational Astrophysics (C3A) at Zhejiang University and chair professor in the School of Physics and Institute for Advanced Study in Physics of Zhejiang University. He received his PhD in astrophysics from Princeton University in 1990. He played a major role in the construction of the theories of the Lyman alpha forest and the cosmic missing baryons. His current active research areas include the escape of Lyman continuum, feedback processes on galaxy formation, cosmological reionization, formation of globular clusters, formation of supermassive black hole seeds, and interstellar turbulence and star formation.


\listoffigures
\listoftables

\end{spacing}
\end{document}